\documentclass[twocolumn,twocolappendix,times]{aastex701}

\usepackage{graphicx}
\usepackage[FIGTOPCAP]{subfigure}
\usepackage{amsmath,booktabs,threeparttable,url, bm}
\usepackage{CJKutf8}
\usepackage{csvsimple}

\newcommand{\cntext}[1]{\begin{CJK*}{UTF8}{bsmi}#1\end{CJK*}}

\newcommand{\change}[1]{{{#1}}}
\newcommand{\Rtwo}[1]{{{#1}}}
\usepackage{soul}

\usepackage{natbib}
\received{\today}

\shorttitle{$r$-process feedback in NS merger outflows}
\shortauthors{Ma et al.}

\begin{document}  
\title{$r$-process Heating Feedback on Disk Outflows from Neutron Star Mergers}

\newcommand*{\NTHUP}{Department of Physics, National Tsing Hua University, Hsinchu 30013, Taiwan}
\newcommand*{\NTHUA}{Institute of Astronomy, National Tsing Hua University, Hsinchu 30013, Taiwan}
\newcommand*{\ASIoP}{Institute of Physics, Academia Sinica, Taipei 115201, Taiwan}
\newcommand*{\ASIAA}{Institute of Astronomy and Astrophysics, Academia Sinica, Taipei 106319, Taiwan}
\newcommand*{\NCTS}{Physics Division, National Center for Theoretical Sciences, Taipei 106319, Taiwan}
\newcommand*{\UoAP}{Department of Physics, University of Alberta, Edmonton, AB T6G 2E1, Canada}

\author[0009-0000-0674-7592]{Li-Ting Ma (\cntext{馬麗婷})}
\affiliation{\NTHUP} \affiliation{\NTHUA} 
\email{liting\_ma@gapp.nthu.edu.tw}

\author[0000-0002-1473-9880]{Kuo-Chuan Pan (\cntext{潘國全})}
\affiliation{\NTHUP} \affiliation{\NTHUA} 
\email{}

\author[0000-0003-4960-8706]{Meng-Ru Wu (\cntext{吳孟儒})}
\affiliation{\ASIoP} \affiliation{\ASIAA} \affiliation{\NCTS}
\email{}

\author[0000-0003-4619-339X]{Rodrigo Fern\'andez}
\affiliation{\UoAP}
\email{}


\begin{abstract}

Neutron star mergers produce $r$-process elements, with yields that are sensitive to the kinematic and thermodynamic properties of the ejecta. These ejecta properties are potentially affected by dynamically-important feedback from $r$-process heating, which is usually not coupled to the hydrodynamics in post-merger simulations modeling the ejecta launching and expansion. The multi-messenger detection of GW170817 showed the importance of producing reliable ejecta predictions, to maximize the diagnostic potential of future events. In this paper, we develop a prescription for including $r$-process heating as a source term in the hydrodynamic equations. This prescription depends on local fluid properties and on the $Y_{e}$ history as recorded by dedicated tracer particles, which exchange information with the grid using the Cloud-in-Cell method. The method is implemented in long-term viscous hydrodynamic simulations of accretion disk outflows to investigate its feedback on ejecta properties. We find that $r$-process heating can increase the unbound disk ejecta mass by $\sim 10\%$ relative to a baseline case that only considers alpha particle recombination. Nuclear heating also enhances the radial velocity of the ejecta with $Y_e < 0.25$ by up to a factor of two, while concurrently suppressing marginally-bound convective ejecta. 
\end{abstract}

\keywords{
\uat{$r$-process}{324}, 
\uat{Neutron stars}{1108}, \uat{Hydrodynamical simulations}{767}}

\section{INTRODUCTION}
Kilonovae are transients powered by the radioactive decay of heavy nuclei synthesized by the rapid neutron-capture process ($r$-process) in ejecta from binary neutron star mergers (BNSM) or black hole-neutron star (BH-NS) mergers. For the first BNSM event detected in gravitational waves by Advanced LIGO and Virgo, GW170817~\citep{PhysRevLett.119.161101}, the electromagnetic counterparts have also been observed from radio to gamma-rays~\citep{Abbott_2017,Alexander_2017,Chornock_2017,Hallinan_2017,McCully_2017,Savchenko_2017,Ruan_2018,Balasubramanian_2021,Hajela_2022}.
For the kilonova counterpart, several follow-up optical and infrared searches recorded detailed photometric and spectroscopic data for $\sim 10$~days after the merger, though late-time detections in certain bands continued for weeks to months \citep{Abbott_2017,Kasen_2017,Pian_2017}. 
Most theoretical kilonova models suggest that GW170817 BNSM ejecta possibly consists of two major components -- a lanthanide-poor component dominating the earlier phase of the blue emission, and a lanthanide-rich component accounting for the later transition to the infrared (e.g. \citealt{McCully_2017,Kasen_2017,Metzger_2019,Margutti_2021}). 

Current numerical simulations of BNSMs have found several mass ejection mechanisms in the dynamical phase of the merger, as well as during the secular evolution of the post-merger remnants (e.g. \citealt{1999A&A...341..499R,Dessart_2008,Bauswein_2013,2013PhRvD..87b4001H,2013MNRAS.435..502F,Shibata_2017,Nedora_2019,Shibata_2019}.) 
Predicting properties relevant to kilonova modeling, such as the ejecta mass, velocity and the nucleosynthesis yields still depends on various physical inputs that are currently not fully known.
These include the equation of state that determines the post-merger lifetime of the hypermassive neutron star (HMNS) (e.g. \citealt{Bauswein_2013,2013PhRvD..87b4001H,lucca_2020,kawaguchi_2023}), the treatment of viscosity arising from the turbulent magnetic fields (e.g. \citealt{Fernandez_2019,Kiuchi_2018,Siegel_2018,Ruiz_2019,Kawaguchi_2022}), the modeling of neutrino-matter interaction and neutrino flavor oscillations (e.g. \citealt{Radice_2016,Joy_2022,Li_2021,Just_2022,PhysRevD.106.103003,Foucart_2023,espino_2023}), or the potential ejecta-ejecta or jet-ejecta interaction (e.g. \citealt{Geng_2019,Hamidani_2019,Nativi_2020,Gottlieb_2022}), among others.  
While current predictions from BNSM simulations are broadly consistent with observations \citep{Metzger_2019,Margutti_2021}, much more efforts remain for quantitatively accurate kilonova modeling.

Among the various ejecta components, the post-merger disk outflow is expected to dominate given the inferred parameters for GW170817 (e.g., \citealt{2017PhRvD..96l3012S}).
One of the key remaining uncertainties in predicting ejecta from disk outflows is the impact of $r$-process heating.
Most long-term post-merger (magneto-)hydrodynamical simulations include energy release from nuclear recombination of alpha particles ($^4$He) and/or from heavier nuclei whose abundances are 
determined by the nuclear statistical equilibrium (NSE), which works together with viscous heating driving the expansion of the disk and the associated mass outflow~\citep{2013MNRAS.435..502F,Just_2015,Kawaguchi_2021}. 
Using a large number ($\sim 4,000$) of heavy nuclei with temperature-dependent partition functions in NSE can result in an additional 1.7 MeV per nucleon released relative to using
only one representative heavy nucleus, which can have a non-negligible impact on the outflow velocity \citep{Fernandez_2024}.
However, this treatment does not take into account the additional energy release due to the $r$-process when non-equilibrium nuclear reactions drive further changes in nuclear compositions when the temperature drops below $\sim 6$~GK. 
While including the energy release due to nuclear transmutations beyond NSE by directly coupling a full nuclear reaction network with multidimensional (magneto-)hydrodynamical simulations is beyond reach, several works have attempted to address the potential impact of additional nuclear heating on properties of different ejecta components from BNSMs and BH-NS mergers as well as the fallback material~\citep{Metzger_2010,Rosswog_2014,Fernandez_2014,Just_2015,Wu_2016,Desai_2019,Ishizaki_2021,Klion_2021,Foucart_2021,Darbha_2021,Kawaguchi_2021,Sneppen_2023,Kawaguchi_2024,Magistrelli_2024,Just_2025}. 
These works have adopted very different ways of approximating the nuclear heating as an additional source term. 
They found different levels of impact due to nuclear heating beyond NSE on ejecta mass and velocity, and their spatial distributions that can possibly affect the detailed modeling of kilonova light curves~\citep{Rosswog_2014,Just_2015,Wu_2016,Klion_2021,Foucart_2021,Darbha_2021,Kawaguchi_2021,Sneppen_2023,Kawaguchi_2024,Magistrelli_2024}, as well as the amount and timescale of fallback matter that may be linked to the observation of $\gamma$-ray bursts \citep{Metzger_2010,Fernandez_2014,Desai_2019,Ishizaki_2021,Foucart_2021}, highlighting the need of incorporating $r$-process heating in simulations.

Particularly for the disk outflows, earlier investigations in \cite{Just_2015} and \cite{Wu_2016} implemented an additional nuclear heating rate that depends only on the local temperature and the sign of the radial velocity of the fluid in axisymmetric, viscous hydrodynamic simulations for BH-disk winds. 
While the temperature-dependent heating rates adopted in \cite{Just_2015} and \cite{Wu_2016} are based on post-processed nuclear reaction network calculations of tracer particles from simulations without including $r$-process heating, they neglected the strong dependence of heating rates on nuclear composition. 
The impact of additional nuclear heating on ejecta mass was found to be potentially significant in \cite{Wu_2016} but negligible in \cite{Just_2015}, possibly related to the different amount of included additional heating at $T\gtrsim 2-3$~GK. 
\cite{Klion_2021} implemented a simple time-dependent heating parametrization, also assumed to be composition-independent, in long-term hydrodynamical simulations mapped from a three-dimensional general relativistic magneto-hydrodynamical BH-disk simulation conducted for 10~s. 
They found that $r$-process heating increases the ejecta velocity by about a factor of 2 when reaching the phase of homologous expansion at $\sim 60$~s, which results in a brighter and bluer kilonova. 
\change{Similar to \cite{Klion_2021}, \cite{Kawaguchi_2021} included r-process heating in long-term hydrodynamical simulations for disk outflows mapped from numerical relativity simulations of BNSMs with massive neutron star remnants. In a related study, \cite{Kawaguchi_2024} performed analogous simulations but for BH-NS mergers, where the central remnant is a black hole. Both studies partially took into account the composition and ejection time dependence of r-process heating in long-term simulations by including the nuclear heating rates obtained from post-processed nucleosynthesis calculations based on tracer particles from numerical relativity simulations. They found relatively minor impacts on ejecta mass and velocity.}

More recently, \cite{Magistrelli_2024} couples a full $r$-process nucleosynthesis reaction network to ejecta represented by Lagrangian particles assumed to expand radially in a ray-by-ray radiation hydrodynamic simulation mapped from a numerical relativity simulation after the central remnant collapses to a BH in $\sim 10$~ms. 
Although the ejecta do not include the long-term secular disk winds in a self-consistent way, they found that the exact nucleosynthesis yields of certain isotopes can be affected by up to a factor of 10 compared to results obtained from post-processed nucleosynthesis calculations. 

In \cite{Just_2025}, they trained a neural network based on post-processed nucleosynthesis calculation data capable of taking several key input quantities including mass density and its time-derivative, temperature, and a few important composition-related parameters to predict the local nuclear heating rate. 
They demonstrated with spherical steady-state winds, BNSM dynamical ejecta, and disk outflows from post-merger NS-disk and BH-disk systems that the neural network captures well the overall nuclear energy release obtained from the corresponding post-processed nuclear reaction network calculations. 
They also found that among the considered BNSM ejecta, nuclear heating makes the largest impact on BH-disk outflows due to its lower velocity --- the ejecta mass and velocity are enhanced by a factor of $\sim 20\%$ and $\sim 50\%$, respectively.

In this study, we also aim to investigate the potential impact of $r$-process heating on the evolution of the post-merger BH-disk system and its outflow properties by including its composition and spatial/temporal dependence. 
We assume that the composition dependence of the nuclear heating rate mainly depends on the electron fraction ($Y_e$) at $T\simeq 6$~GK.  We utilize a large number of ``memory'' tracer particles, an externally-tabulated $Y_e$ dependent heating rate, and the technique of Cloud-in-Cell (CIC) method \citep{1969JCoPh...3..494B} to distribute the heating rate information carried by the memory particles to the simulation domain.
In Section~\ref{sec:method}, we describe the numerical methods adopted for our hydrodynamic simulations, including the developed heating method. In Section~\ref{sec:OverallEvol}, we present the overall evolution observed in our simulations. Sections~\ref{sec:TracerResult}, \ref{sec:HeatingResult} and \ref{sec:UniformHeating} discuss the main results, focusing on the details of $r$-process heating feedback. Finally, we summarize our findings in Section~\ref{sec:conclusion}.

\section{METHOD AND MODELS}\label{sec:method}
\subsection{Hydrodynamic Simulation} \label{sec:hydro}
We use the FLASH code, version 3.2 \citep{2000ApJS..131..273F,dubey2009}, to perform viscous hydrodynamic simulations with the setup described in \citet{2013MNRAS.435..502F,2014MNRAS.441.3444M,FKMQ14}, and \citet{PhysRevD.106.103003}. The simulations employ two-dimensional axisymmetric spherical coordinates. The computational domain extends radially from $\sim$ 8.8~km to $8.8 \times 10^5$~km, logarithmically divided into 640 cells, resulting in a cell size of $\Delta r \sim 1.49$~km for the innermost cell, with outflow boundaries at both ends. In the polar direction, the angle $\theta$ spans from 0 to $\pi$, distributed evenly in terms of $\cos{\theta}$ across 112 cells, employing reflecting boundaries. The finest resolution at the equator corresponds to $\Delta \theta \sim 0.018$~radian. The code employs domain decomposition to parallelize execution using the Message Passing Interface (e.g., \citealt{MPI}), with each processor working with a domain block of equal number of cells.

\change{The initial condition is set to match the inferred parameters of the event GW170817 with a post-merger BH remnant and an accretion disk \citep{PhysRevLett.119.161101,PhysRevD.106.103003}}. At the center of the simulation box, we place a spinning BH with a mass of 2.65~$M_\odot$ and spin parameter of $0.8$ as the remnant compact object after the merger binary, which acts on its surroundings through a pseudo-Newtonian gravitational potential \citep{1996ApJ...461..565A}. The simulations are initialized from an equilibrium torus characterized by constant specific angular momentum, a constant entropy of 8~$k_{\rm B}$/baryon, and an electron fraction ($Y_{e}$) of 0.1. The disk has a total mass of 0.1~$M_\odot$, with the maximum density occurring at a radius of 50 km. We implement  an $\alpha$-disk model, having an azimuthal viscous stress with a viscosity coefficient $\alpha = 0.03$ or 0.06, which quantifies the strength of angular momentum transport \citep{1973A&A....24..337S}. 

In our simulations, we use the Helmholtz equation of state \citep{2000ApJS..126..501T}, with protons, neutrons, and alpha particles in nuclear statistical equilibrium. The nuclear binding energy of alpha particles is included in the internal energy. A 3-species leakage scheme with emission and annular lightbulb-type absorption is employed for the neutrino transport, with the updates described in \citet{PhysRevD.106.103003}. Passive tracer particles are used to trace the fluid properties throughout the whole simulation. In each simulation, we deploy $10^4$ tracers with equal effective mass, initially distributed according to the density profile (cf. \citealt{Wu_2016}).

\subsection{Implementation of $r$-process heating feedback} \label{sec:heating}
\begin{figure}
    \epsscale{1.2}
	\plotone{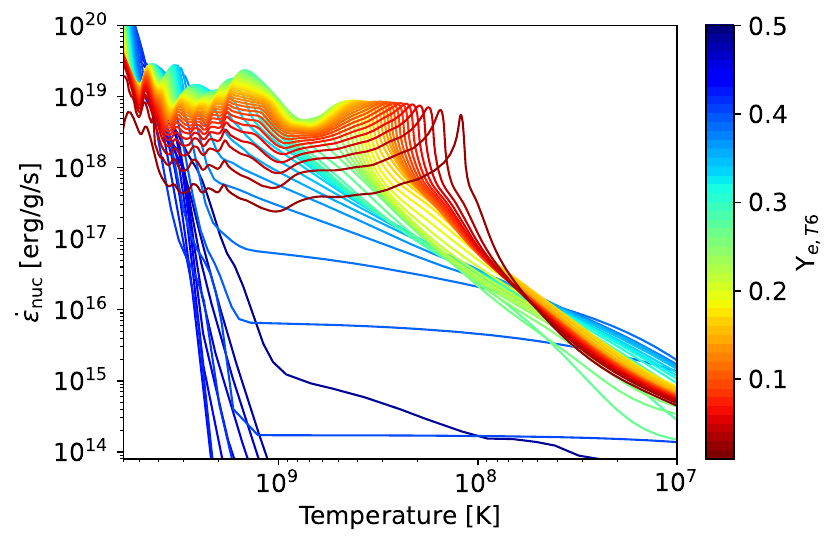}
	\caption{\label{fig_Qdot}
	Specific nuclear energy release rates $\dot\varepsilon_{\rm nuc}$ as functions of temperature for 50 sample parametrized expansion trajectories with different initial $Y_e$ (see text for details). Our heating prescription interpolates the heating rate based on temperature and $Y_{e}$, using the instantaneous fluid temperature (as a proxy for time) and two trajectories whose $Y_{e, T6}$ (i.e. the $Y_{e}$ value when it first reached temperature 6~GK) bracket the value that the fluid element has.}
\end{figure}

To achieve the goal of implementing the composition-dependent nuclear heating rate during $r$-process, we adopt the rate of nuclear energy release profiles from nucleosynthesis calculations performed over a parametrized expansion trajectory with an assumed entropy per nucleon of $30$~$k_{\rm B}$ and an expansion time scale $\tau=10$~ms \citep{Lippuner_2015,PhysRevLett.122.062701} for 50 different initial $Y_e$ (at 8~GK) values uniformly spaced from 0.01 to 0.50. 
\change{The specific parameters, $s = 30$~$k_{\rm B}$ per nucleon and $\tau = 10$~ms, were chosen because the $\rho(T)$ curve resulting from this profile is representative of the thermal trajectories observed in the simulated outflow tracers.}
Figure~\ref{fig_Qdot} shows the nuclear energy release rates as functions of temperature for all 50 different initial $Y_e$, highlighting extended (shortened) nuclear energy release for lower (higher) $Y_e$ values. 
For simplicity, we assume that 50\% of the released nuclear energy is carried away by neutrinos, with the remaining half contributing to the fluid's internal energy as extra heating source \citep{Metzger_2010_406,Barnes_2016}.



Since the hydrodynamical simulations are performed on Eulerian grids while the heating rates are associated with Lagrangian fluid particles, we take the following approach to map the heating rate data onto the simulation grid. We incorporate $9.9 \times 10^5$ additional tracers in models that consider $r$-process heating.
These additional tracers, which we refer to as ``memory tracers'', are designed to record the fluid elements' $Y_{e}$ history. Within the 12 domain blocks covering the initial equilibrium torus, 82500 memory tracers are uniformly initialized in each block, aiming to maximize the coverage across all cells (See Appendix~\ref{appx:tracer} for details on the convergence test determining the necessary number of memory tracers.)

At each time step, we map the fluid temperature and $Y_{e}$ onto each tracer. When a tracer passes through the temperature of 6~GK for the first time, its $Y_{e}$ is recorded and designated as the initial $Y_{e}$ in our heating profile; we denote this value as $Y_{e,T6}$. Heating is applied to the outflow material in our fiducial models only when the local fluid temperature drops below a threshold temperature ($T_{\rm heat}$) of 4~GK during the simulation, approximating the onset of the $r$-process. To assess the sensitivity of this threshold, we also consider a higher threshold temperature of 6~GK (see Section~\ref{sec:HeatingResult}). 
We apply the same heating rate with a negative sign for inflowing material ($v_{r} < 0$), which serves as cooling. This treatment prevents convecting tracers from contributing redundant heating energy when they flow outward multiple times. 
The specific heating rate is calculated through interpolation based on temperature and $Y_{e,T6}$ for each tracer. The interpolation procedure involves three steps: (1) identifying the closest two tabulated $Y_{e, T6}$ values that bracket the tracer’s $Y_{e, T6}$; (2) performing temperature interpolation in log scale at each of these $Y_{e, T6}$ values using the tracer’s temperature; and (3) interpolating between the two resulting heating rates in linear scale to account for differences in $Y_{e, T6}$. For tracers with $Y_{e, T6} > 0.5$, we adopt the heating rate corresponding to $Y_{e, T6} = 0.5$ to avoid extrapolation beyond the boundary of the table. The computed specific heating rate is then subsequently mapped back onto the grid-based fluid elements.
Besides taking different threshold temperature, we also present a resolution test for the adopted number of $Y_{e}$ bins used for heating rate interpolation by introducing a low-resolution model, al03T4Ye5, which uses 5 bins instead of 50. See Appendix~\ref{appx:Ye5} for detail discussion.

Throughout this study, we employ the Cloud-in-Cell (CIC) method \citep{1969JCoPh...3..494B} to map information between grid cells and tracers. In the mesh-to-particle mapping, field quantities such as temperature and $Y_e$ are assigned to each tracer using a first-order distance-weighted interpolation based on the values of the cell containing the tracer and its adjacent cells. Conversely, in the particle-to-mesh mapping, tracer properties (the specific heating rate) are distributed to the surrounding grid cells using the same distance-weighted scheme, where each tracer’s contribution to a given cell is proportional to its distance to the cell center. This method ensures smooth transitions across cell boundaries and allows a limited number of tracers to exchange information with a larger set of grid cells.

\subsection{Models}\label{sec:model}
We explore ten models with two different viscosities and two temperature thresholds for applying heating; all the models are listed in Table~\ref{tab1}. We choose the viscosity parameter $\alpha = 0.03$ as our fiducial model, and $\alpha = 0.06$ for comparison (labeled al03 and al06, respectively). We perform simulations with and without $r$-process heating (the latter denoted by nH in the model name in Table~\ref{tab1}) for both viscosity values, to investigate the heating effect. Five of our models adopt 4~GK as the threshold temperature for heating (named with T4). Model al03T6 and al03T6nC use a higher threshold temperature $T_{\rm heat} = 6$~GK to check sensitivity of results to this parameter.

For the model al03K, we adopt the spatially-uniform heating rate described by a four-segment broken power-law function of time given by Equation~3 of \cite{Klion_2021}.
Compared to our fiducial heating rate implementation, we will show in Section~\ref{sec:UniformHeating} that this prescription may have overestimated (underestimated) heating in early (late) phase and will discuss the consequence on the resulting impact of heating. 

Additionally, we performed three supplementary models, denoted as nC, in which cooling is disabled for the inflow materials. This modification enables us to assess the contribution of inflow to the overall dynamics.
All models that include $r$-process heating evolve up to 20~s, except for model al03T4, which evolves up to 16.6~s. Models without heating (al03nH and al06nH) are performed for a longer duration of 40~s.

Matter from the grid or tracer particles are considered unbound when their Bernoulli parameter satisfies $B_{e} > 0$ and reach radial distances \change{$ \ge 10^{3}$~km}. The Bernoulli parameter is defined as:
\begin{align}\label{eq:unbound}
  B_{e} = \dfrac{1}{2} \mathbf{v}^2 + e_{int} + \dfrac{P}{\rho} + \Phi
\end{align}
where $\mathbf{v}$ is the total fluid velocity (including rotation), $e_{int}$ the specific internal energy, $P$ the total pressure, $\rho$ the mass density, and $\Phi$ the gravitational potential. 

\begin{table*}
  \caption{\label{tab1}
 Models investigated. The first five columns show the model name and their set-up parameters; from left to right are the model names, viscosity parameter $\alpha$, whether r heating is included, threshold temperature for heating $T_{\rm heat}$, and the resolution for interpolation in $Y_{e, T6}$. The following three columns are the resulting unbound mass $M_{ej}$, mass averaged radial velocity at $T = 1$~GK $\langle v_r \rangle _{T9}$ and $r$-process heating energy $Q_{r}$. See Section~\ref{sec:model} for the definition of unbound. The last two columns are the ratio of the mass averaged radial velocity at $T = 1$~GK $\langle v_r \rangle _{T9}$ compared to the corresponding values in the no-heating (nH) models, we separate the tracers in two groups, $\mathcal{R}_{\langle v_r \rangle;< 0.25}$ for low-$Y_{e, T6}$ tracers, and $\mathcal{R}_{\langle v_r \rangle;\geq 0.25}$ for high-$Y_{e, T6}$ tracers.}
  \centering
  \begin{tabular}{cccccccccc}
  \hline \hline
  Model & $\alpha$  & heating  & $T_{\rm heat}$ & $Y_{e, T6}$ bins &
  $M_{ej}$ [$M_{\odot}$] & $\langle v_r \rangle _{T9}$ [$10^{-2}$c] &
  $Q_{r}$ [$10^{50}$erg] & $\mathcal{R}_{\langle v_r \rangle;< 0.25}$ &
  $\mathcal{R}_{\langle v_r \rangle;\geq 0.25}$\\
  \hline 
  al03nH & 0.03 & no   & --  & -- &  0.027 & 2.85 & 0     & 1.00 & 1.00\\
  al06nH & 0.06 & no   & --  & -- &  0.030 & 3.62 & 0     & 1.00 & 1.00\\
  al03T4 & 0.03 & yes  & 4GK & 50 &  0.030 & 5.49 & 3.20  & 2.09 & 1.89\\
  al03T6 & 0.03 & yes  & 6GK & 50 &  0.031 & 5.99 & 4.18  & 2.78 & 1.90\\
  al06T4 & 0.06 & yes  & 4GK & 50 &  0.033 & 6.13 & 3.10  & 2.25 & 1.44\\
  al03K  & 0.03 & yes  & --  & -- &  0.031 & 4.45 & 2.56  & 2.28 & 1.33\\
  \hline
  al03T4Ye5 & 0.03 & yes  & 4GK & 5  &  0.030 & 4.88 & 2.59  & 1.96 & 1.65\\
  al03T4nC  & 0.03 & yes  & 4GK & 50 &  0.030 & 5.71 & 3.38  & 2.35 & 1.91\\
  al03T6nC  & 0.03 & yes  & 6GK & 50 &  0.032 & 6.11 & 4.36  & 2.96 & 1.94\\
  al06T4nC  & 0.06 & yes  & 4GK & 50 &  0.032 & 6.04 & 3.19  & 2.04 & 1.49\\
  \hline \hline
  \end{tabular}
\end{table*}

\section{Overall evolution}\label{sec:OverallEvol}
\begin{figure*}
    \centering
    \includegraphics[width=1.0\textwidth]{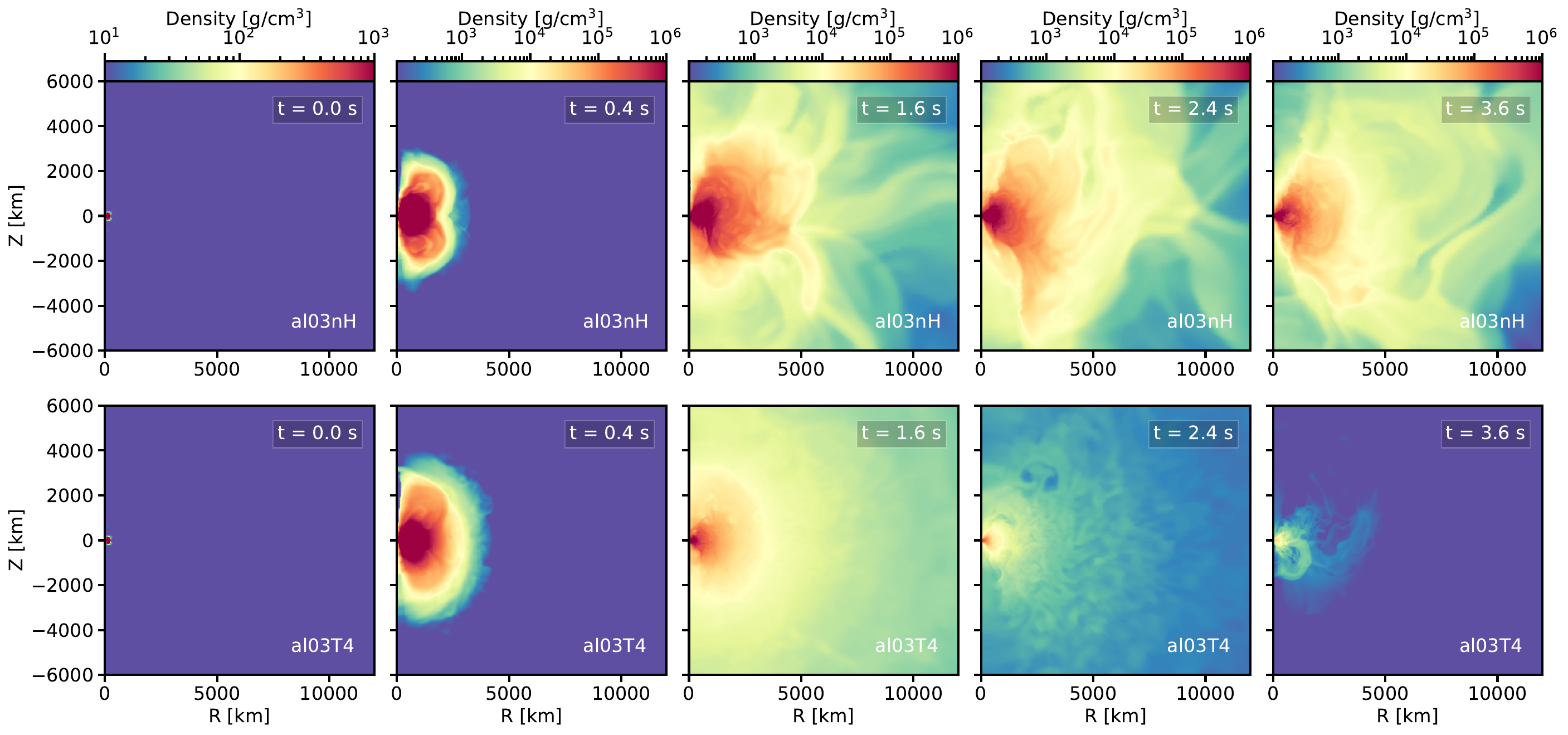} 
    \caption{\label{fig_slice}
    Snapshots of density in our fiducial model at various times, as labeled, comparing a baseline case with no $r$-process heating (top row, model al03nH) and one that includes our heating rate prescription with default settings (bottom row, model al03T4). Both models share the color map at the same time. The model with nuclear heating produces an outflow with more spherical morphology and on a shorter timescale than the model without $r$-process heating.}
\end{figure*}
Figure~\ref{fig_slice} illustrates the density evolution of models al03nH and al03T4 from $t = 0$ (left) to $t = 3.6$~s (right). We start our simulation with an equilibrium torus, where gravitational forces are balanced by centrifugal force and gas pressure. As the disk evolves, angular momentum is transported from the inner to the outer disk. This redistribution weakens the centrifugal force in the inner disk, allowing a significant portion of mass to accrete onto the central BH via gravity. Meanwhile, the outer disk gains angular momentum and expands outward. In the inner region, neutrino cooling also takes away thermal energy, further enhancing the accretion.

\begin{figure*}
	\plotone{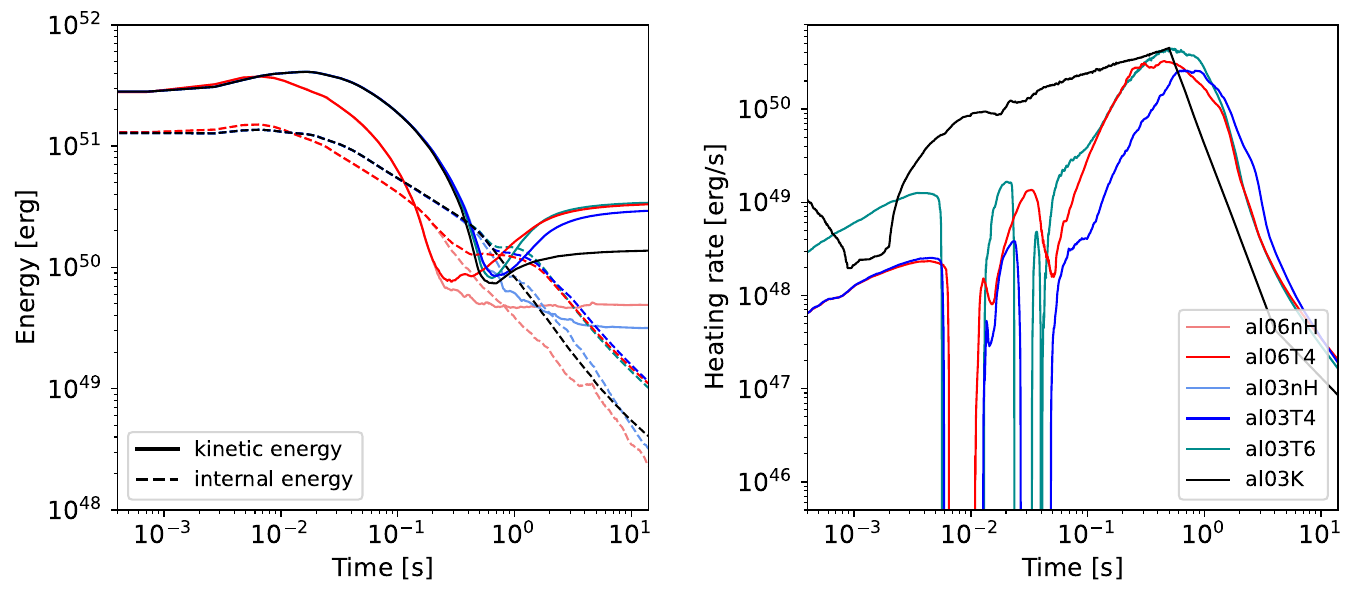}
	\caption{\label{fig_evol}
    \emph{Left:} energy evolution for selected models. Solid and dashed lines represent the total kinetic and internal energy in the simulation domain for each model, respectively. \emph{Right:} total heating rates as functions of time for various models in different colors, as indicated in the legend.}
\end{figure*}

The accretion in the inner region and the expansion at the outer region lead to a decrease in density, which in turn reduces the efficiency of neutrino cooling on the viscous timescale ($\sim 300$\,ms for $\alpha=0.03$). As neutrino cooling becomes insufficient to offset viscous heating, gas pressure once again supports the inner disk against gravity, thus slowing down the accretion rate. At the same time, the remaining material in the outer region gains energy from viscous heating and nuclear recombination, thus accelerating its expansion. The launching of this outflow occurs within approximately $t \leq 1$~s and results in a decrease in both temperature and density of the outflowing matter \citep{2013MNRAS.435..502F}. 
As the internal energy of the outermost material is gradually converted into kinetic energy, the kinetic energy rises again between 0.2 and 1 s and becomes the dominant component of the system’s energy budget.  This transition is presented in the left panel of Figure~\ref{fig_evol}, which depicts the energy evolution across seven models. Solid lines represent the fluid's total kinetic energy, while dashed lines represent the total internal energy. The different line colors denote various models, as indicated in the legend. 

Figure~\ref{fig_slice} shows that the model which includes nuclear heating launches the outflow on a shorter timescale. By $t = 3.6$~s, only a small amount of mass remains in model al03T4. In contrast, the model without heating (al03nH) retains a significant amount of mass, indicating a slower and less efficient outflow. The model with $r$-process heating also produces a more spherical outflow, particularly evident at $t = 1.6$~s. 
\Rtwo{The global increase in thermal energy drives this spherical morphology through rapid, isotropic expansion, which smooths large scale asymmetries and makes radial velocity dominant over tangential convective fluctuation.}


\section{Tracer evolution and $r$-process heating feedback}\label{sec:TracerResult}

\begin{figure*}
	\plotone{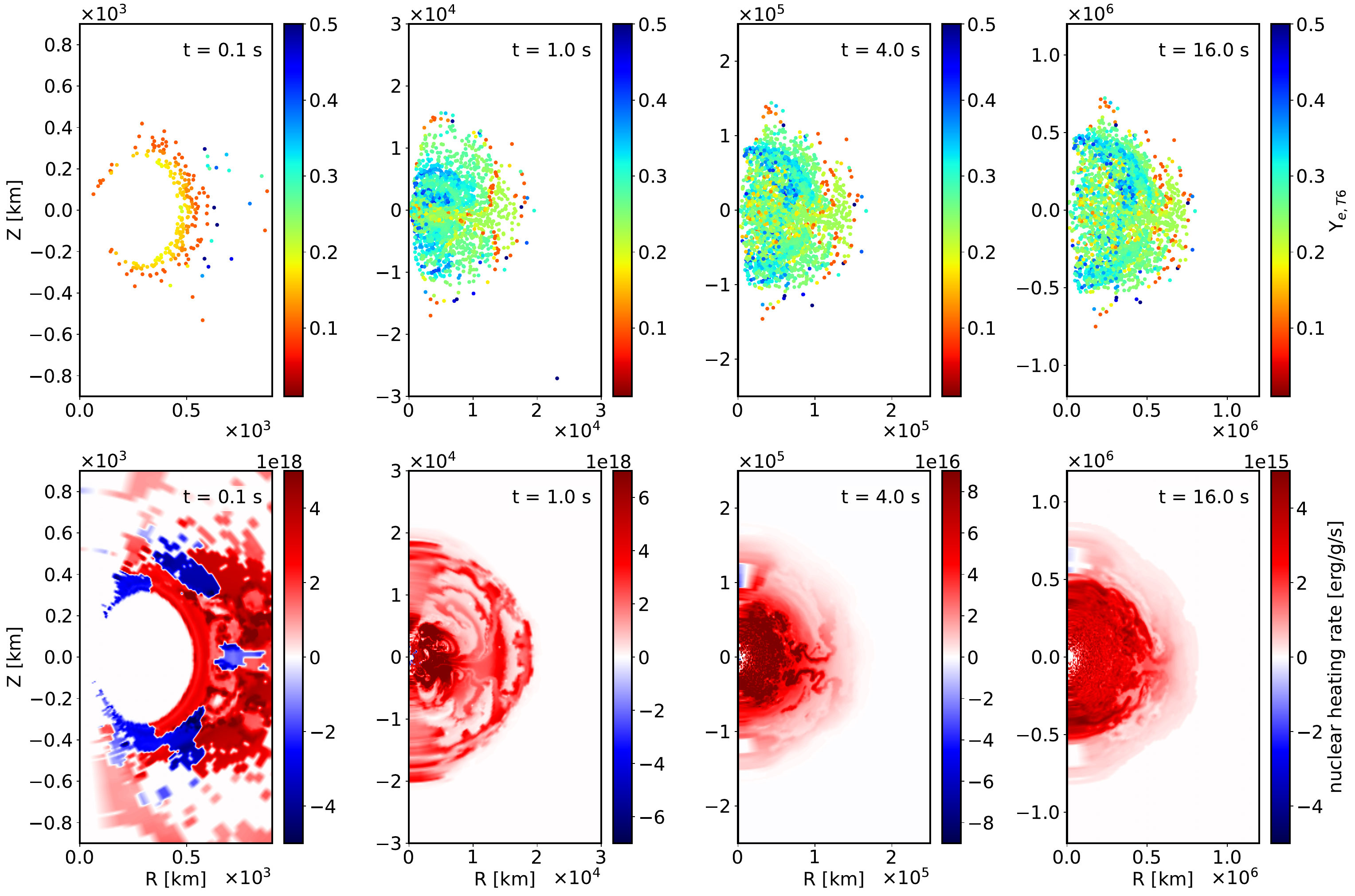}
	\caption{\label{fig_yet6_HR}
	The upper four panels depict the distribution of tracers from 100~ms to 16~s (left to right) \change{for model al03T4}. The color map represents each tracer's $Y_{e, T6}$ value. The empty region in the left panel (at t = 0.1~s) corresponds to the initial torus, where temperature exceeds 6~GK. In this high-temperature region, tracers are not yet assigned with their $Y_{e, T6}$ value. Heating rate is also not required under such conditions. The lower panels display the specific nuclear heating rate derived from tracers corresponding to the upper panels. Heating regions (outflows) are shown in red, while cooling regions (inflows) are in blue.}
\end{figure*}

To better understand the heating implementation, we track the evolution of passive tracers in our simulations, which are initially distributed according to the density profile, and analyze their behavior. Initially, tracers originating from the torus outer regions rapidly reach the critical temperature threshold of $T=6$~GK while maintaining its low $Y_e$. As shown in the upper panels of Figure~\ref{fig_yet6_HR}, the value of $Y_{e, T6}$ of these early, fast-moving tracers \change{in model al03T4} remains low. This condition leads to high nuclear heating rates (Figure~\ref{fig_Qdot}) and an earlier onset of heating contributions. The lower panels of Figure~\ref{fig_yet6_HR} show that a strong heating rate (indicated in red) first appears in those regions corresponding to the locations of low-$Y_e$ tracers identified in the upper panel. These low-$Y_e$ tracers contribute to the early energy input into the system and are efficiently accelerated outward. 

\begin{figure}
    \epsscale{1.2}
	\plotone{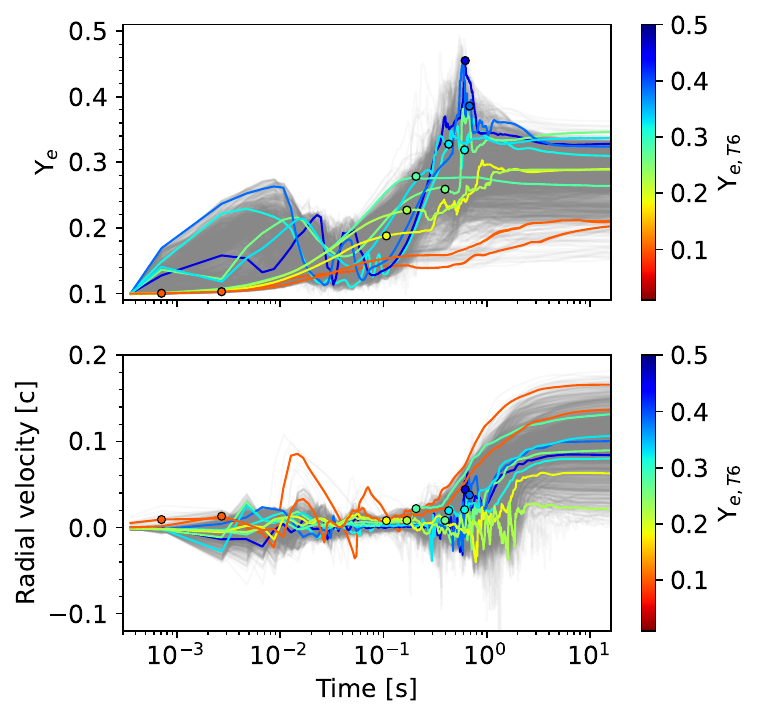}
	\caption{\label{fig_Ye_vr}
	Time evolution of the $Y_{e}$ (upper panel) and radial velocity (lower panel) for model al03T4. Grey lines in the background represent all unbound tracers. 10 representative unbound tracers are highlighted in color, selected to cover a range of $Y_{e,T6}$ values, with colors assigned according to the color map. Dotted points mark the time when each tracer first passes through $T = 6$~GK, indicating the onset of its contribution to nuclear heating or cooling.}
\end{figure}

Figure~\ref{fig_Ye_vr} further quantifies this behavior; in the lower panel, tracers with lower $Y_{e, T6}$ (colored towards red) experience a substantial boost in radial velocity. The upper panel confirms that their $Y_{e}$ value remains low (only slightly increased) since weak interactions have only slightly altered their $Y_{e}$ for a short time.
In contrast, tracers originating from the inner disk remain longer in high-density and high-temperature regions, where weak interactions gradually raise their $Y_e$, as depicted in Figure~\ref{fig_yet6_HR} (upper panel). These inner tracers cross the $T=6$~GK threshold later, resulting in a delayed heating contribution and generally lower heating rates, given their higher $Y_{e, T6}$ values (Figure~\ref{fig_Qdot}). 

We take the radial velocity when the tracer passes through $T = 1$~GK for the last time as a representative radial velocity at the end of nucleosynthesis for each tracer. Since the density-weighted distributing tracers have the same effective mass, we calculate the mass-averaged radial velocity for all the unbound tracers in each model and list them as $\langle v_r \rangle _{T9}$ in Table~\ref{tab1}. To understand the behavior of the different components, we separate the unbound tracers into two groups, the low-$Y_{e, T6}$ component for tracers with  $Y_{e, T6} < 0.25$ and high-$Y_{e, T6}$ component for tracers with $Y_{e, T6} \geq 0.25$. For these two components, we compute the mass-averaged radial velocity and calculate the ratio for the models with heating to the corresponding model without heating, which is listed as $\mathcal{R}_{\langle v_r \rangle;< 0.25}$ and $\mathcal{R}_{\langle v_r \rangle;\geq 0.25}$ in Table~\ref{tab1}. The radial velocity of low-$Y_{e, T6}$ material is enhanced by more than a factor of two when including $r$-process heating feedback.

When considering the combined contribution of all the tracers, the overall heating rate distribution is shown in the lower panel of Figure~\ref{fig_yet6_HR}. Initially, the heating volume concentrates near the inner regions surrounding the original torus and the heating rate intensifies later as more tracers reach the temperature threshold. The heating rate then declines over time as material moves out, which is consistent with the heating profile as a function of (decreasing) temperature we applied (Figure~\ref{fig_Qdot}).

Despite the local variations in heating rates associated with the tracers' composition, the global morphology of the ejecta remains nearly spherical. 
\Rtwo{Convection and turbulence redistribute tracers across different latitudes, producing spatially uniform heating. This uniform heating deposited into the ejecta, increases the thermal pressure and drives rapid expansion, thereby shaping the spherical ejecta morphology. }
In Figure~\ref{fig_yet6_HR}, the upper panels show the spatial mixing of tracer components, and the lower panels show how the heating distribution becomes isotropic over time. 

\begin{figure}
    \epsscale{1.2}
	\plotone{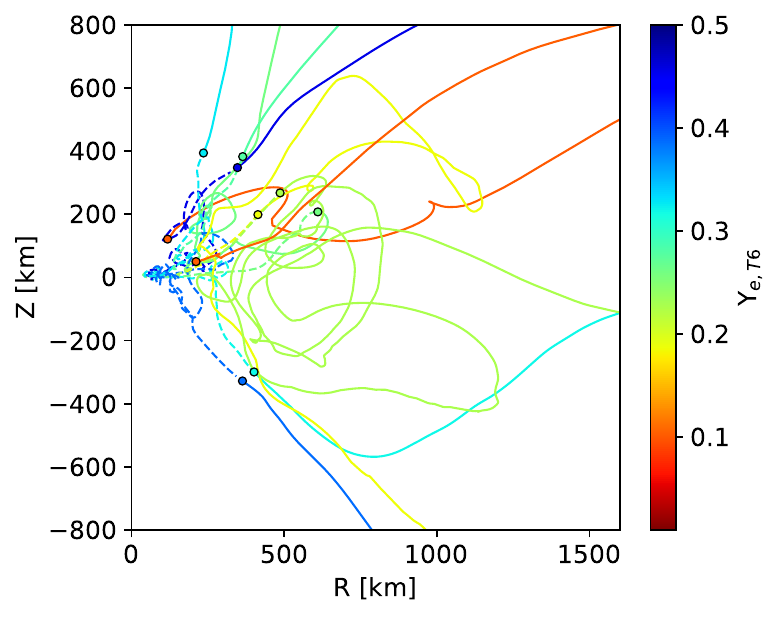}
	\caption{\label{fig_traj}
    Trajectory paths from model al03T4 are displayed, featuring the same 10 selected tracers as in Figure~\ref{fig_Ye_vr}, with colors indicating their $Y_{e,T6}$ values. Filled circles mark the time when each tracer first passes through $T = 6$~GK. Dashed and solid lines depict the trajectory segments before and after the onset of contribution to nuclear heating or cooling.
}
\end{figure}

To further understand this phenomenon, we illustrate the spatial paths of selected tracers in Figure~\ref{fig_traj}. Circles mark the point at which each tracer first passes through $T=6$~GK; the dashed segments indicate tracer trajectories before reaching 6~GK, while the solid segments show trajectories after 6~GK. The highly intertwined trajectories indicate significant convective motion that mixes material from various regions across the torus, which suppresses directional asymmetries and causes a globally spherical ejecta morphology, albeit local spatial non-uniformities in heating rate persist.

\section{Interplay of viscous heating and $r$-process heating}\label{sec:HeatingResult}

Figure~\ref{fig_evol} illustrates the temporal evolution of energy components and heating rates for the models listed in Table~\ref{tab1}. The solid and dashed lines in the left panel represent the total kinetic and internal energy, respectively. The right panel shows the total heating rates as functions of time for various models in different colors, as indicated in the legend.

During the initial phase of the simulation (within hundreds of milliseconds), both internal and kinetic energy decrease due to accretion, as described in Section~\ref{sec:OverallEvol}.
Models with $\alpha = 0.06$ evolve more rapidly than those with $\alpha = 0.03$, as reflected in the earlier decline of both internal and kinetic energies. This faster evolution is a consequence of a more efficient energy transport associated with higher viscosity parameter. High $\alpha$ also causes faster conversion of thermal energy into kinetic energy and an earlier rise in the kinetic energy at around $t = 0.2$~s.

In the early stage of evolution (before $5 \times 10^{-3}$~s), the heating rate for model al03T6 \change{(green line)} is approximately six times greater than that of al03T4 and al06T4 \change{(blue and red lines)}. This difference is due to the heating onset threshold being set at a temperature of 6 GK rather than 4 GK, enabling earlier heating in model al03T6. Across all models, the total heating rate becomes negative between $5 \times 10^{-3}$ and $4 \times 10^{-2}$~s due to notable inward matter flow, which induces cooling as described in Section~\ref{sec:heating}. Subsequently, the total heating rate becomes positive again and peaks between 4 and 10~s depending on the model, followed by a decline consistent with the specific heating rate profiles presented in Figure~\ref{fig_Qdot}.

\begin{figure*}
	\plotone{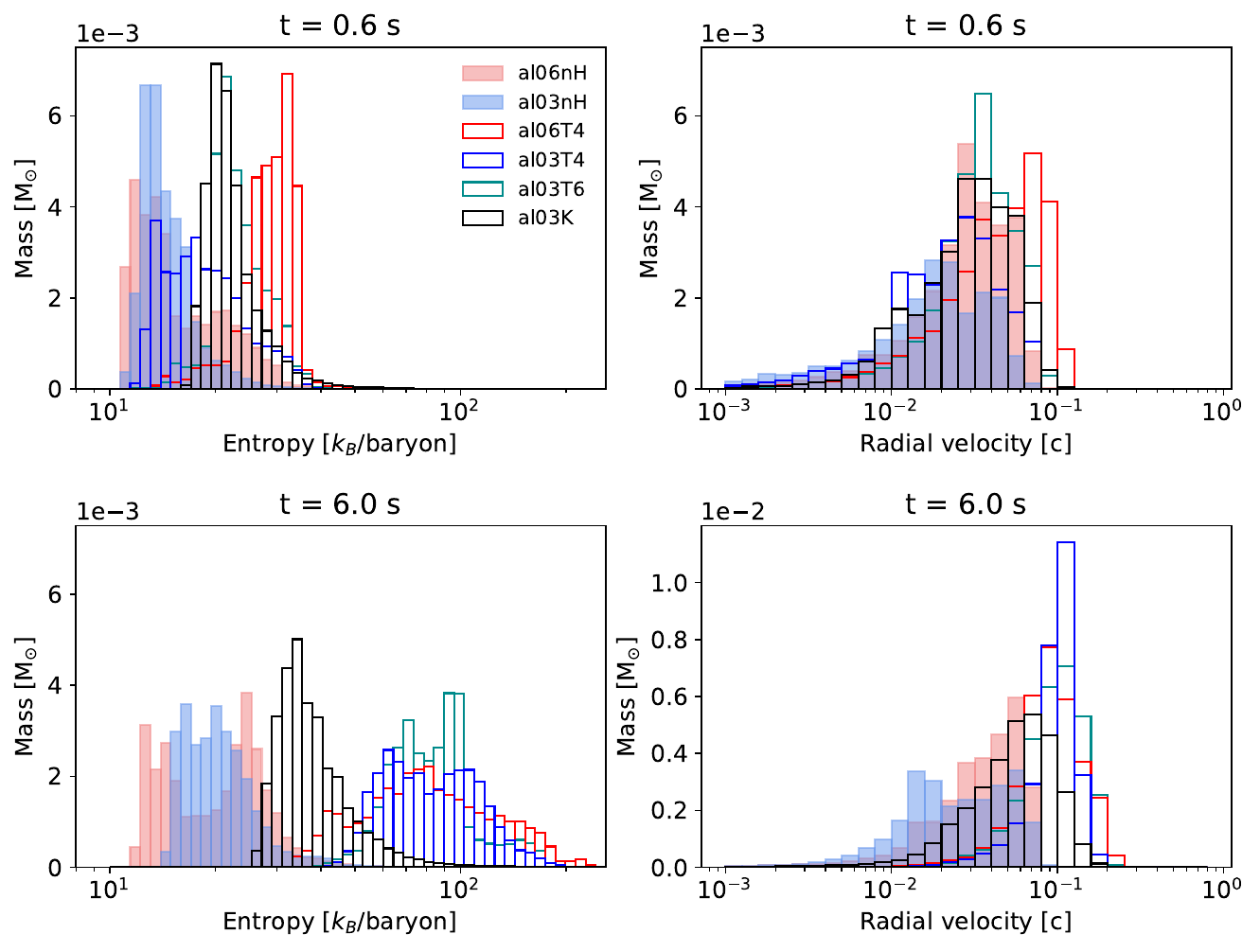}
	\caption{\label{fig_hist}
	Mass distribution in different entropy and velocity bins. The upper two panels show the mass distribution at 600~ms and the bottom two show the result at t = 6~s.}
\end{figure*}

Figure~\ref{fig_hist} presents mass distributions of unbound ejecta, computed using grid data over the entire simulation domain, as functions of entropy (left panels) and radial velocity (right panels) at two different times: $t = 0.6$ (top panels) and $t = 6.0 $ (bottom panels). The histograms refer to different models, as indicated in the legend. Models without heating (with heating) are represented by filled (hollow) boxes.

The radial velocity distributions at $t= 0.6$\,s indicate maximum ejecta velocities of approximately 0.1~c, with most models peaking around 0.02~c and an extended tail toward lower velocities. Models with $\alpha = 0.06$ exhibit faster ejecta, the peak velocities are $\sim 0.02$~c and $\sim 0.1$~c for models without and with heating \change{(light red filled bars and red outlined bars)}, respectively. These models expand more rapidly into the outer, lower-temperature regions. The upper panels of Figure~\ref{fig_hist} illustrate that viscosity dominates over $r$-process heating during the early phase (within the first few hundred milliseconds).

By t = 6~s, the entropy distribution in models without nuclear heating (al03nH and al06nH, \change{blue and red filled bars}) shows only minor changes relative to that at $t=0.6$\,s. In contrast, models that include $r$-process heating display a five-fold increase in entropy, caused by the additional energy input from nuclear heating. The radial velocity distributions indicate an overall increase, peaking at approximately 0.15~c, highlighting the acceleration of the ejecta due to $r$-process heating feedback. Models al06T4 and al03T6 \change{(red and green outlined bars)} even achieve maximum radial velocities of 0.25~c. 
Velocity distributions of models without heating only slightly increase relative to $t=0.6$\,s due to thermal energy conversion into kinetic energy. Over time, the heating rate deposits energy into the fluid, and on a timescale of seconds, heating alters the thermodynamic properties of the ejecta. The results in Figure~\ref{fig_hist} suggest that while viscous heating governs the early dynamical evolution, $r$-process heating becomes the dominant factor after a few seconds.

\begin{figure}
    \epsscale{1.2}
	\plotone{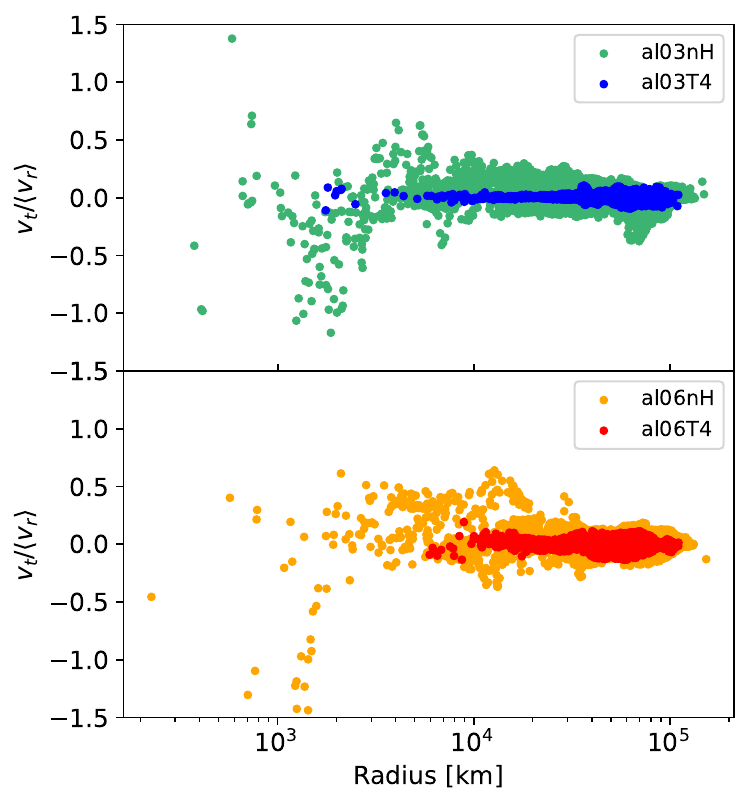}
	\caption{\label{fig_vt}
    \change{Ratio of the tangential velocity to the angle-averaged radial velocity for unbound tracers as a function of radius. Tracers from models with and without nuclear heating are shown in different colors, as indicated in the legend. The upper and lower panels correspond to two viscosity parameters, $\alpha = 0.03$ and $\alpha = 0.06$, respectively. Snapshots are selected for each model at times corresponding to similar positions of the forward shock front, ensuring a consistent comparison of the outflow structure.}}
\end{figure}

Figure~\ref{fig_vt} shows the \change{ratio of tangential velocity to angle-averaged radial velocity $v_{t}/\langle v_{r}\rangle$ of unbound tracers at t = 6~s for models without heating (al03nH and al06nH), t = 2.4~s for model al03T4, and t = 2.8~s for model al06T4.} 
In the inner region, $r \lesssim 10^{5}$~km, tracers from models without nuclear heating (al03nH and al06nH, \change{green and orange dots}) exhibit \change{significant dispersion in $v_{t}/\langle v_{r}\rangle$ around zero, indicating strong convective activity.} 
In contrast, for models that include heating \change{(blue and red dots), $v_{t}/\langle v_{r}\rangle$ remains closer to zero in this region}, indicating that convection is suppressed. 
\change{However, this suppression primarily affects convection near the marginally-bound ejecta close to the black hole, and convective and turbulent motions remain active on larger scales, driving the overall spherical morphology of the ejecta.} This behavior is consistent with the low-velocity end of the radial velocity histograms in the lower panel of Figure~\ref{fig_hist}, where the models without heating contain a greater amount of low-velocity ejecta than the heating models at t = 6~s. These slow-moving ejecta reflect convective motion near the BH. 

\begin{figure}
    \epsscale{1.2}
	\plotone{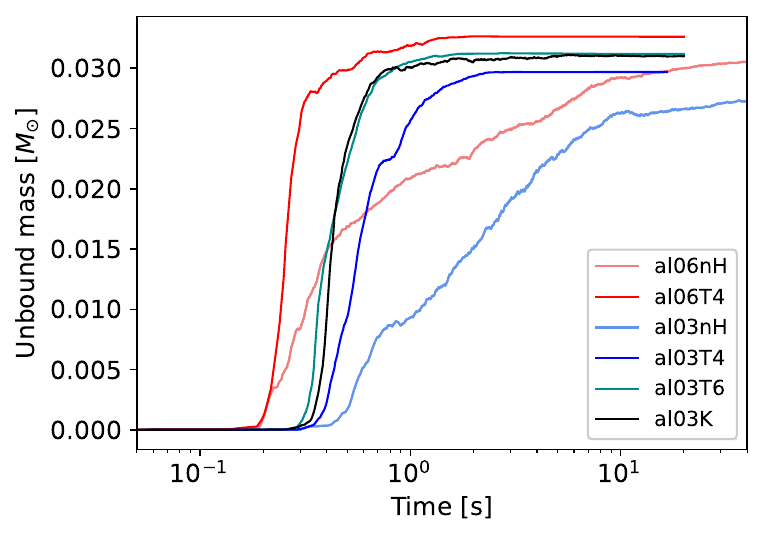}
	\caption{\label{fig_unbound}
	Time evolution of the unbound mass reaching \change{$r = 10^3$~km}, as defined in Equation~(\ref{eq:unbound}), for various models. Each model is represented by a different color, as indicated in the legend.}
\end{figure}

We analyze tracers that are initially distributed with the fluid density. Following the gravitational binding criterion defined in Section~\ref{sec:model}, Figure~\ref{fig_unbound} shows the temporal evolution of the cumulative unbound mass reaching \change{$r = 10^3$~km} for different models, as indicated in the legend. Models that include heating exhibit rapid unbound mass growth, reaching saturation within $\sim 0.5$~s. 
In contrast, non-heating (nH) models \change{(light blue and light red lines)} gradually increase over roughly 2~s. This points out that neglecting $r$-process heating results in slower ejection of material, with the saturation timescale differing by 1.5~s. The final unbound mass values ($M_{ej}$) are listed in Table~\ref{tab1}, showing a 10\% increase in models that include heating. Additionally, models with higher viscosity ($\alpha = 0.06$) evolve more quickly and eject 10\% more mass than those with $\alpha = 0.03$, consistent with the expectation that enhanced angular momentum transport facilitates earlier mass ejection.

\section{Comparison with time-dependent uniform heating rate}\label{sec:UniformHeating}
To assess the importance of composition and outflow history dependence of $r$-process heating feedback, we perform an additional simulation al03K with the time-dependent $r$-process heating rate adopted by \cite{Klion_2021}, which approximates the heating rate as a four-segment broken power-law function of time that is independent of the nuclear composition, i.e., spatially uniform. 
All other physical parameters are identical to those in our fiducial model.

The time evolution of the heating rate in model al03K is shown on the right panel of Figure~\ref{fig_evol} (black line). Compared to our composition-dependent models, al03K exhibits a higher heating rate during the early phase ($t < 0.5$~s). Since the heating function from \cite{Klion_2021} is purely time-dependent and independent of composition, the spatial variation in the heating rate arises solely from the mass distribution. After 0.5~s, the outflow rate becomes stable, and the heating rate follows the broken power-law as the adopted function. However, the magnitude of heating in this later phase is lower than that in our composition-dependent models. When integrated over time, the total heating energy $Q_r$ in model al03K is significantly lower than in the composition-dependent cases (see Table~\ref{tab1}), resulting in a final kinetic energy nearly two times smaller, as shown in the left panel of Figure~\ref{fig_evol}. 
The total ejecta internal energy is similar to that of the model without heating (al03nH and al06nH), which differs significantly from other models that include heating.  The time evolution of both internal and kinetic energy in al03K are almost the same as that of model al03T6 before 0.6~s. Consequently, the total unbound mass in al03K is identical to al03T6. Based on this comparison, we infer that the ejecta mass is primarily determined by heating at early times, while heating at later times mainly influences the asymptotic velocity. As the ejecta is already unbound, the subsequent heating energy contributes to further accelerating the expansion. This trend is also evident in the left panels of Figure~\ref{fig_hist}. At $t = 0.6$~s, the entropy distribution of model al03K closely resembles al03T6. However, by $t = 6$~s, the entropy in al03K increases by only a factor of two, which is significantly smaller than in the other heating models. This comparatively minor entropy increase demonstrates that al03K receives less $Q_r$ than the composition-dependent heating models.

Despite the overall smaller energy, heating in model al03K still affects the fluid properties. As shown in Figure~\ref{fig_hist}, the radial velocities are lower than in the composition-dependent heating models but still significantly higher than in the no-heating case. This indicates that although the spatially uniform heating prescription can accelerate the outflow (consistent with findings from \citealt{Klion_2021}), a uniform heating treatment can underestimate the effect on the feedback of $r$-process heating.

\section{DISCUSSION}\label{discussion}

Recently, \citet{Just_2025} have reported a new method for coupling r-process heating
with the hydrodynamics that relies on machine-learning techniques, including an application of
the method to a BH torus system. Their findings are broadly consistent with ours: (1) their BH torus with r-process heating ejects more mass and on a shorter timescale relative to the case without heating; (2) r-process heating has the highest impact on the lower end of the outflow velocity distribution, removing slower (marginally bound) material; and (3) the ejecta morphology is more spherical in the case with heating relative to the case with no heating. 
Quantitatively, \citet{Just_2025} find an increase in BH torus ejecta mass and average velocity of $20\%$ and $50\%$, respectively, when including r-process heating. This compares with our finding of $10\%$ increase in ejecta mass and $70\%-90\%$ increase in ejecta velocity when including r-process heating in our fiducial models.
This quantitative difference can be attributed to a few factors: (1) the average velocities of
our baseline models without heating are lower than that of their model without heating, thus
the velocity boost for a comparable nuclear energy gain should be larger, 
(2) we report velocities when particles last reach a temperature of $1$\,GK instead of at the end of the simulation, and (3) we report  unbound ejecta mass with a specific unbinding criterion. Given the drastically different
approaches, we consider overall agreement with \citet{Just_2025} to be excellent.

Angular momentum transport in realistic merger disks is expected to occur via magnetohydrodynamic (MHD) turbulence driven by the magnetorotational instability (e.g., \citealt{balbus_1998}) instead of shear viscosity. Previous long-term 3D MHD simulations of BH accretion disks from NS mergers have shown that such turbulence combined with large-scale magnetic stresses can drive faster and more massive disk winds for strong poloidal fields, with typical average outflow velocities around 0.1~c instead of 0.03\,c for viscous hydrodynamic models
\citep{Siegel_2018,Fernandez_2019,miller_2019,Just_2022,kiuchi_2023}. 
In this regime of higher baseline velocities, the relative impact of r-process heating is likely to be reduced, leading to more moderate velocity increases of the order of $\sim 10\%$, if the energy gain per baryon remains constant, 
compared to the factor of two enhancement observed in our simulations.
For toroidal fields or weak poloidal fields, average outflow velocities are closer to the viscous hydrodynamic values \citep{Christie_2019,Fahlman_2022,hayashi_2023}, 
in which case $r$-process heating can still be an order unity correction to the outflow velocities.
The field geometry immediately after the merger is complex, with a mixture of poloidal and toroidal components spanning a range of spatial scales due to turbulence (e.g., \citealt{most_2021,aguilera_2023,izquierdo_2024,kiuchi_2024}), 
hence the definitive impact of r-process heating on outflow velocities with realistic field geometries remains an area of active study.

Our current heating prescription interpolates the nuclear energy release based on $Y_{e, T6}$ and the instantaneous temperature. However, outflows from merger disks exhibit a broad range of entropies and expansion timescales, which also affect nucleosynthesis and heating rates. \change{Within this framework, we expect that variations in the adopted heating parameters may lead to changes in the total nuclear energy release within a factor of three.} Expanding the heating tables to include these additional parameters in future work can improve the physical accuracy of the heating rate.

Another study worth exploring is extending our simulation into three dimensions. A primary concern for extending the simulation into three dimensions is the computational cost of implementing our tracer-based heating approach. In two dimensions, our current method relies on $9.9 \times 10^5$ memory tracers to achieve sufficient spatial coverage, already resulting in a significant increase in computational expense compared to the baseline model, by roughly a factor of six, primarily due to communication between the grid and particles. A direct extension to three dimensions would require orders of magnitude more tracers to maintain comparable resolution, which is likely impractical. However, since the large-scale outflow remains broadly axisymmetric (e.g., \citealt{Christie_2019}), one potential solution is to adopt an axisymmetric heating rate. This approach can keep computational costs comparable to the two-dimensional case while preserving the composition-dependent heating prescription.

\section{CONCLUSION}\label{sec:conclusion}
This study investigates the impact of $r$-process heating feedback on the dynamics and properties of disk outflows from post-merger BNS systems. 
To address the heating feedback, we developed a novel prescription for $r$-process heating, in which the heating rate varies with the local fluid temperature and $Y_e$ history. We implement our heating prescription on long-term, two-dimensional viscous hydrodynamic simulations. We performed simulations with different viscosity parameters, heating threshold temperatures, $Y_e$ resolutions, and cooling treatments for inflowing material. For comparison, we also included a model employing a spatially uniform, time-dependent heating prescription based on previous literature.

Our results demonstrate that $r$-process heating significantly influences the disk outflows' dynamical evolution and final properties. We found that $r$-process heating increases the unbound disk ejecta mass by approximately 10\% compared to models without heating. Our results also show a substantial enhancement in the radial velocity of neutron-rich outflows ($Y_e < 0.25$), with velocities a factor of two higher than in models without heating. This acceleration is attributed to the early and intense heating in low-$Y_e$ regions, which drives faster expansion. In contrast, the higher-$Y_e$ component experiences delayed and weaker heating owing to longer residence times in the inner disk region.
We have also shown that using a time-dependent heating rate that neglects the dependence on outflow history and nuclear composition underpredicts the outflow velocity.
 
We tested the sensitivity of our results on two numerical free parameters: the temperature threshold for the onset of heating ($T_{\rm heat}$) and the $Y_{e,T6}$ resolution used for heating rate interpolation. $T_{\rm heat}$ mainly altered the total heating energy, moderately affecting our outcomes. The qualitative trends are robust for different $Y_{e,T6}$ resolutions. Model al03T4Ye5 shows the same unbound mass but slightly lower ejecta velocity than al03T4. Additionally, we found that disabling cooling for inflows has only a minimal effect on the results.

This work emphasizes the importance of including composition-dependent $r$-process heating in post-merger simulations of NS merger disk outflows. Accurate modeling of this feedback mechanism is essential for reliable predictions of kilonova lightcurves and the interpretation of future multi-messenger observations. Nucleosynthesis calculation and radiative transfer modeling are required for deriving the associated observables, such as abundance yields and kilonova lightcurves, which we leave for future work.

\begin{acknowledgments}
 We thank Oliver Just, Gabriel Mart\'inez-Pinedo, Albert Sneppen, and Zewei Xiong for helpful discussions. This work is supported by the National Science and Technology Council of Taiwan through grants 113-2112-M-007-031, 114-2112-M-007-020, by the Center for Informatics and Computation in Astronomy (CICA) at National Tsing Hua University through a grant from the Ministry of Education of Taiwan. 
 MRW acknowledges support of the National Science and Technology Council, Taiwan under Grant No.~111-2628-M-001-003-MY4, the Academia Sinica under Project No.~AS-IV-114-M04, and the Physics Division of the National Center for Theoretical Sciences, Taiwan. 
 RF acknowledges support from the Natural Sciences and Engineering Research
Council of Canada (NSERC) through Discovery Grant RGPIN-2022-03463, and is also grateful for
the hospitality of the Institute of Physics, Academia Sinica, where part of this work was conducted.
 {\tt FLASH} was in part developed by the DOE NNSA-ASC OASCR Flash Center at the University of Chicago. The simulations and data analysis have been carried out on the CICA cluster at National Tsing Hua University.
\end{acknowledgments}

\bibliography{reference}{}

\appendix 

\section{Evaluation of memory tracer coverage}\label{appx:tracer}
In our simulations, memory tracers provide the heating rate information required for interpolation throughout the heating domain. 
Initially, tracers are uniformly distributed within 12 blocks, each consisting of $64 \times 28$ cells. Within each block, tracers are placed uniformly and randomly across the cells to sample the fluid properties. We choose a uniform initial distribution rather than a density-weighted one to maximize coverage across all cells. Since density varies with the grid cells, density weighting would concentrate tracers in high-density regions and leave low-density regions undersampled. 

As the simulation evolves, tracers advect with the fluid flow, resulting in non-uniform coverage over time. Maintaining complete coverage of all cells becomes challenging, and using an excessively large number of tracers is impractical due to the computational cost. Therefore, we perform a set of test simulations to determine the minimum number of tracers necessary to achieve sufficient coverage while maintaining affordable computational time.

We tested eight different tracer population sizes per block: $2 \times 10^2$, $1 \times 10^3$, $2 \times 10^3$, $2 \times 10^4$, $4 \times 10^4$, $8.25 \times 10^4$, $2 \times 10^5$, and $2 \times 10^6$ tracers. Each configuration evolved for two seconds, including the first second, during which heating rates peak and impact the simulation results most. 

To assess tracer coverage, we identified all the cells satisfying the heating criteria: temperature $< 4$~GK, outward radial velocity, and density exceeding the ambient density floor. We then determined the subset of these cells lacking tracer coverage, where coverage is defined by the availability of tracer information through the particle-to-mesh CIC mapping scheme. A cell is considered uncovered only if it and its four neighboring cells contain no tracers and thus cannot provide heating rate values.

\begin{figure}
    \epsscale{1.2}
	\plotone{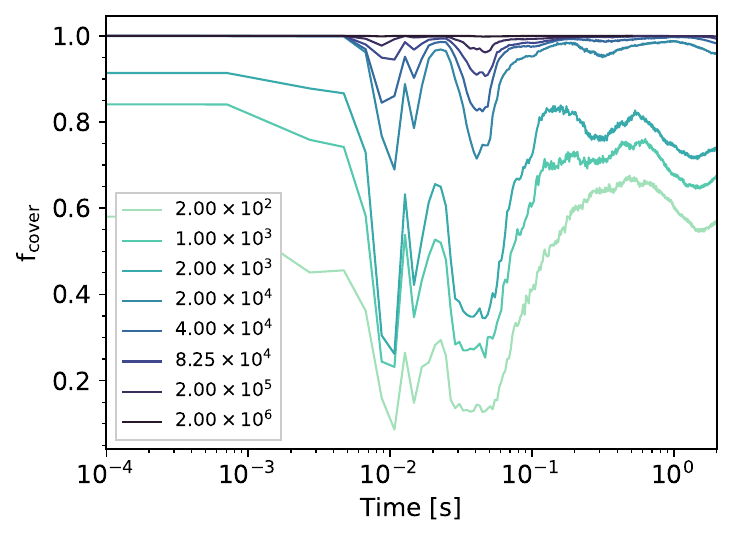}
	\caption{\label{fig_UniformTracer}
    Time evolution of the ratio of total cell mass that both satisfies the heating criteria and covering by tracers to the total cell mass satisfying the heating criteria, $f_{\rm cover}$. Different lines represent different initial numbers of memory tracers assigned per block, as indicated in the legend.}
\end{figure}

By calculating the fraction of the total mass requiring heating and covered by tracers relative to the total mass matching the heating conditions, referred as $f_{\rm cover}$, we can quantify the accuracy and reliability of the tracer sampling approach. As shown in Figure~\ref{fig_UniformTracer}, the tracer coverage improves monotonically with the increasing number of memory tracers. When the number of tracers per block is less than $2 \times 10^3$, tracers fail to cover all the cells even at the beginning of the simulation, resulting in insufficient coverage for accurate heating implementation. The coverage profiles converge for cases with more than $2\times10^4$ tracers per block, except during two specific time intervals around 10~ms and 40~ms, where noticeable dips occur. Notably, for the case of $8.25 \times 10^4$ tracers per block, the first dip near 10~ms significantly reduces compared to cases with fewer tracers. Based on this analysis, we use $8.25 \times 10^4$ tracers per block for our production runs as a practical trade-off between computational cost and coverage accuracy: the coverage remains above 98\% for most of the simulation time, with a temporary decrease to approximately 90\% near 40~ms. However, this drop is short-lived, and coverage quickly recovers thereafter.

\section{Numerical treatments on heating implementation}

\subsection{Resolution of $Y_{e, T6}$ for heating rate interpolation}\label{appx:Ye5}
To assess the impact of $Y_{e, T6}$  resolution on the heating rate interpolation, we compare two models: the fiducial model al03T4, which adopts a high-resolution of 50 bins in $Y_{e, T6}$, and a low-resolution model al03T4Ye5, which uses only five bins. Both models share identical initial conditions and setup configurations, differing only on the resolution of the $Y_{e, T6}$ data points in the nuclear heating rate table.

\begin{figure*}
	\plotone{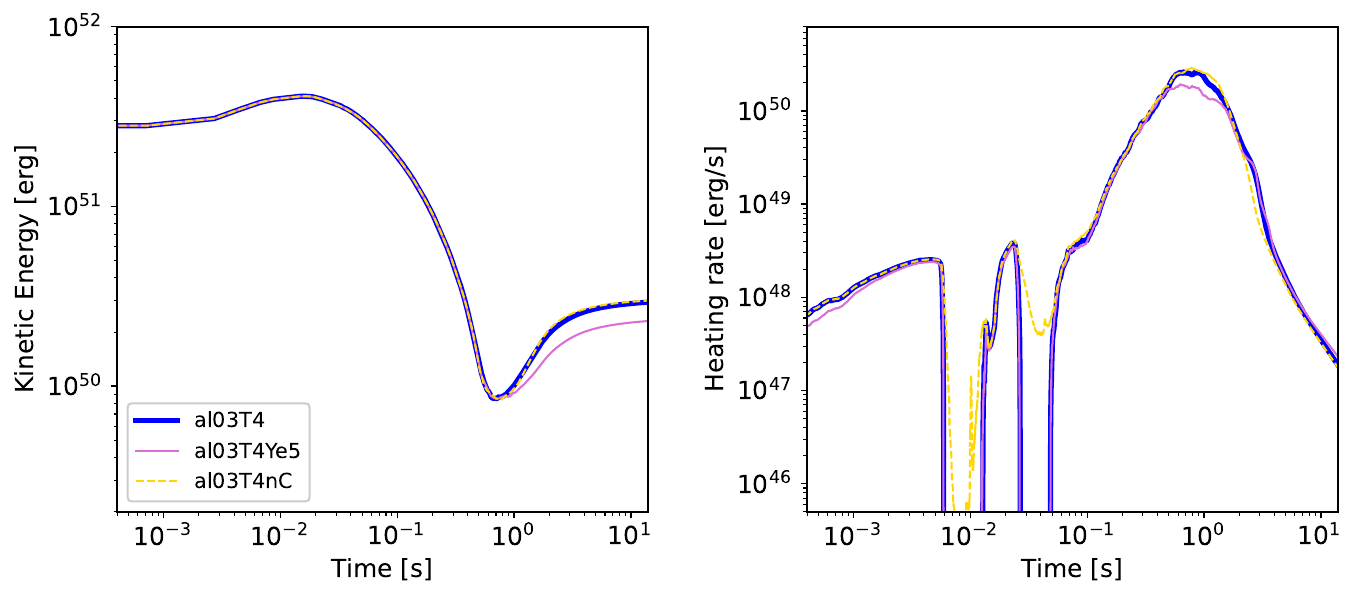}
	\caption{\label{fig_evol_nC}
    The left panel shows the time evolution of kinetic energy, and the right panel shows the time evolution of the total heating rate for model al03T4, al03T4Ye5 and al03T4nC. As indicated in the legend, solid blue line represent simulations using the fiducial heating treatment, while yellow dashed line correspond to models that do not apply cooling for inflowing material. The pink line represents model al03T4Ye5, which uses a lower resolution in $Y_{e, T6}$ for the heating rate calculation.}
\end{figure*}

As shown in the right panel of Figure~\ref{fig_evol_nC}, the total heating rate as a function of time for the low-resolution model (pink) closely follows that of the fiducial case (blue), with only a slight reduction in amplitude. However, this minor difference in heating rate accumulates over time and significantly impacts the final energy. In the left panel, the kinetic energy of model al03T4Ye5 is obviously lower than that of al03T4 despite their similar heating rate profiles. Quantitatively, the total heating energy in al03T4 is approximately 1.2 times greater than the low-resolution model, as presented in Table~\ref{tab1}. This is also reflected in the average radial velocity, which is higher in al03T4 than in al03T4Ye5. 

These results demonstrate that even modest differences in the heating rate profile, arising from coarse interpolation in $Y_{e, T6}$, can affect the final kinetic energy and ejecta velocity. Despite this, the two models are broadly consistent with each other. As their early-time energy evolution is identical, they result in having the same ejecta mass, and the resulting average radial velocity, $v_{r}$, only differs by $\sim 10$ \%.

\subsection{Treatment of cooling for inflowing material}\label{appx:nC}
To evaluate the sensitivity of our results to the assumption regarding energy loss from inflowing material, we perform three additional simulations: al03T4nC, al03T6nC, and al06T4nC, in which the cooling term for inflowing regions is disabled. All other physical and numerical settings remain identical to their corresponding models.

As shown in the right panel of Figure~\ref{fig_evol_nC}, the heating rates of the no-cooling models (yellow dashed line) differ from those of the fiducial models (blue solid line) only around $t \sim 10$~ms. Nevertheless, the total heating rate is two orders of magnitude lower than other times during this period. Outside this time interval, the total heating rate profile remain consistent between two sets of models. Consequently, the difference in heating results is minimal. The kinetic energy profiles in the left panel show only minor differences between the dashed and solid lines. A quantitative comparison in Table~\ref{tab1} shows that the total heating energy and the resulting average radial velocities in the no-cooling cases are slightly higher than those in the fiducial models.

Since the inflowing mass is not the major contributor to the total heating energy, neglecting cooling for inflowing material does not significantly alters the simulation outcomes. This test confirms the robustness of our approach, demonstrating that our overall conclusions are not sensitive to the detail of the treatment of inflow material.

\end{document}